\magnification=1200
\baselineskip=18truept
\input epsf

%preprint or not
\def\preprint{Y}
%draft or not
\def\draftversion{N}
\def\cap{\hsize=4.6in}

\if \draftversion Y


\fi

% Figure
\def\figure#1#2#3{\if \preprint Y \vskip .5cm \midinsert \epsfxsize=#3truein
\centerline{\epsffile{figure_#1_eps}} \halign{##\hfill\quad
&\vtop{\parindent=0pt \hsize=4.6in \strut## \strut}\cr {\bf Figure
#1.}&#2 \cr} \endinsert \fi}

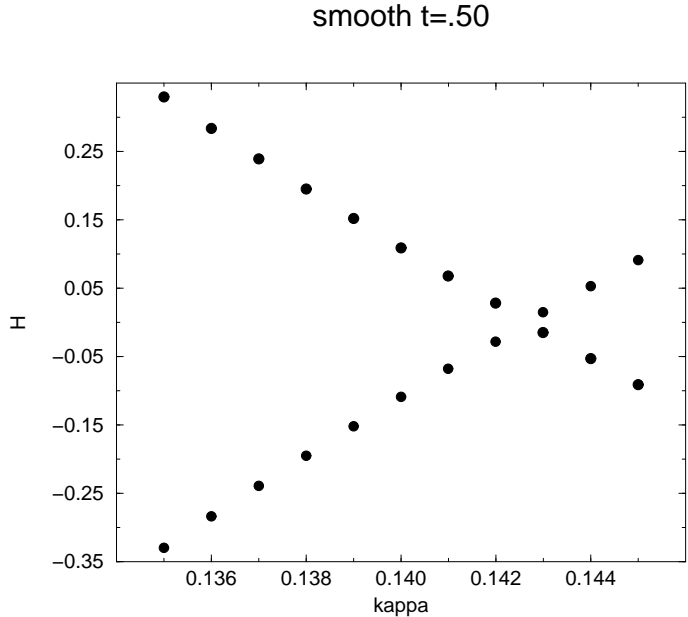
\def\captionone{\cap First cross-section through double cone;
within numerical accuracy: crossing.}
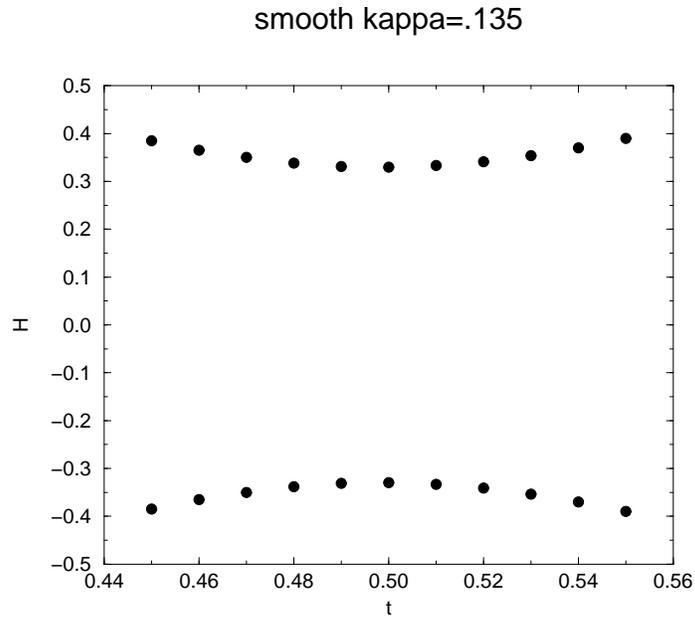
\def\captiontwo{\cap Second cross-section through double cone;
avoided crossing.}
\def\captionthree{\cap Third cross-section through double cone;
avoided crossing.}
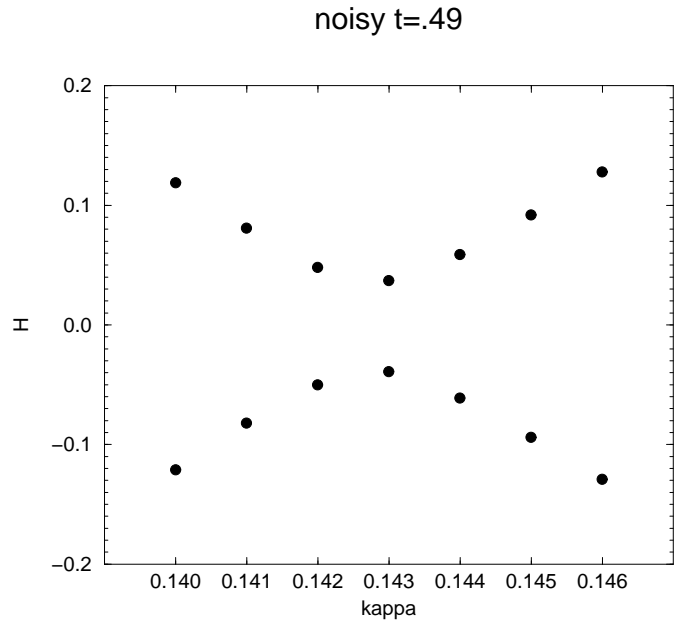
\def\captionfour{\cap First cross-section through double cone;
avoided crossing.}
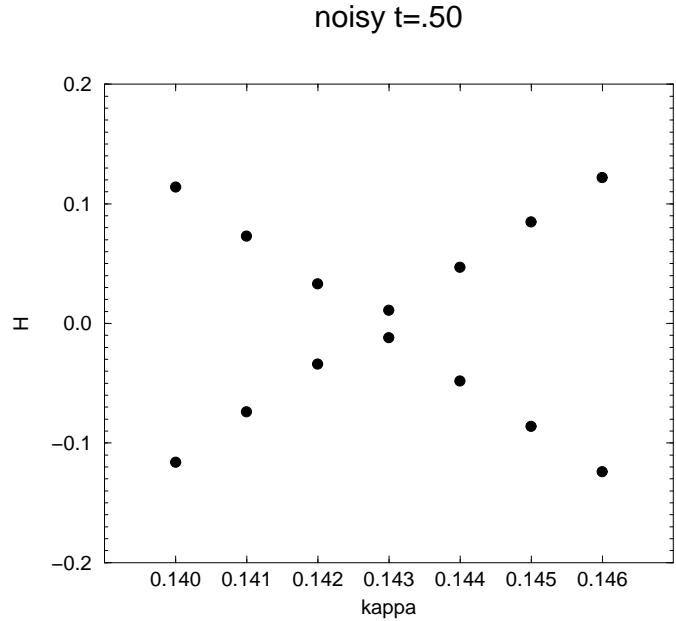
\def\captionfive{\cap Second cross-section through double cone;
within numerical accuracy: crossing.}
\def\captionsix{\cap Third cross-section through double cone;
avoided crossing.}

\def\witten{1}
\def\elitzur{2}
\def\gaumenelson{3}
\def\npblong{4}
\def\daemi{5}
\def\kaplanschmaltz{6}
\def\odddim{7}
\def\geom{8}
\def\sureal{9}
\def\exactlymassless{10}
\def\berry{11}
\def\ritz{12}
\def\urs{13}
\def\GW{14}
\def\hasen{15}
\def\narayanan{16}
\def\luscher{17}
\def\rebbi{18}

\line{\hfill RU--98--22}
\vskip 2cm
\centerline {\bf Witten's $SU(2)$ anomaly on 
the lattice.}
\vskip 1cm
\centerline{Herbert Neuberger}
\vskip .25cm
\centerline{\tt neuberg@physics.rutgers.edu}
\vskip 1.5cm
\centerline{\it Department of Physics and Astronomy}
\centerline{\it Rutgers University}
\centerline{\it Piscataway, NJ 08855--0849}
\vskip 2cm
\centerline{\bf Abstract}
\vskip .5cm
Witten's anomaly for $SU(2)$ with a single 
$I={1\over 2}$ Weyl fermion in four dimension is shown
to be reproduced by the lattice overlap. The mechanism
is based on Berry's phase, and on the analyticity of the
matrix $H$ by which the overlap is defined. 

\vskip .3cm
\vfill
\eject
Although Witten's anomaly (WA) [\witten] 
in $SU(2)$ gauge theory
cannot be {\it directly} seen in perturbation theory,
(see [\elitzur]) it is, conceptually, the simplest
example of a chiral anomaly in four dimensions 
[\gaumenelson]. It seems therefore reasonable to expect
any general {\it non-perturbative} scheme for chiral
gauge theories to contain WA in an 
{\it evident} way. The 
first arguments for the presence of WA in the overlap 
[\npblong,\daemi] (not relying on an embedding of WA
in a perturbative anomaly of a larger group [\elitzur])
were given in [\kaplanschmaltz].  Further work 
related to the realization of 
global anomalies in
the overlap was carried out 
in [\odddim,\geom,\sureal]. I know
of no evidence for WA in other non-perturbative
attempts to construct chiral gauge theories. 
In this paper I follow the 
line of thought
laid down in [\geom] and 
[\sureal] to show that any
smooth phase choice (what is meant by this
will be made clear later on) 
for the overlap necessarily
reproduces WA.

The main object in the overlap 
is the many body fermion 
ground state $|v\{U_l\}>$ which depends parametrically
on the background made out of all $SU(2)$ link
variables $U_l$. The second quantized fermion
system is non-interacting, with hamiltonian $
{\cal H}=a^\dagger H a$ and the dependence on $\{U_l\}$
comes in through the matrix $H$ sandwiched between
the fermionic creation/annihilation operators $a^\dagger
, a$. $H$ is $\gamma_5$ times the standard Wilson-Dirac
operator with the hopping parameter $\kappa$ 
set to $\kappa_0$, 
somewhere in the range ${1\over{2(d-1)}}> \kappa_0
> {1\over{2d}}$ for Euclidean dimension $d$. 
For us,
$d=4$. While the exact structure of $H$ is unimportant,
what follows crucially and equally depends both on the
locality of $H$ and its analyticity in the link
variables $\{U_l\}$.

The chiral determinant is given by the ``overlap''
$<v_{\rm ref}|v\{U_l\}>$ where, in the simplest version,
$|v_{\rm ref}>$ is $U_l$-independent and defined just
as $|v\{U_l\}>$, only 
now $H=\gamma_5$. For the case that
the gauge group is $SU(2)$ and the fermions have 
$I={1\over 2}$ one can choose a basis where $H\{U_l\}$
is real for all $\{U_l\}$. 
An explicit basis was written
down in [\sureal], but the fact that such a basis
exists must have been known to many people.
In this basis it becomes evident that $|v\{U_l\}>$ and
$|v_{\rm ref}>$ can be both taken real. Under a
gauge transformation $U\rightarrow U^g$ we have
${\cal H} \{U^g_l\}=G^\dagger (g) {\cal H}\{U_l\}
G(g)
$ induced by $H\{U^g_l\}=g^\dagger H\{U_l\} g$.
$G(g)$ represents $g$ in the fermionic Fock
space. 
WA simply means that a choice of the states
$|v\{U_l\}>$ that is smooth in the gauge background (and in $\kappa_0$) 
would result in
$$
|v\{U^g_l \}>=-G^\dagger (g) |v\{U_l \}>\eqno{(1)}$$
for a lattice gauge transformation $g$ which is
reasonably close to a continuum gauge transformation
$g$ that is an element of the non-trivial homotopy
class of maps $T^4\rightarrow SU(2)$. Since 
$|v_{\rm ref}>$ obeys $|v_{\rm ref}>=G^\dagger (g)
|v_{\rm ref}>$ for any gauge transformation, the sign 
in equation (1) induces a sign switch in
the regularized chiral determinant of the overlap
as $\{ U_l\}\rightarrow \{U^g_l\}$ 
and hence a violation of gauge invariance. This is
the single kind of gauge violation that can happen 
in the overlap; the square of the overlap is gauge
invariant, and moreover, obeys a simple identity
derived in [\exactlymassless]:
$$
(<v_{\rm ref} | v\{U_l\}>)^2 =\det{{1+V}\over 2},
~~~~~~~~V\equiv\gamma_5\epsilon (H),\eqno{(2)}$$
where $\epsilon$ is the sign function. 

Following [\witten] I 
consider a path parameterized
by $0<t<1$ connecting a configuration $\{U_{l,0}\}$
to $\{U^g_{l,0}\}$ for a nontrivial continuum
$g$, suitably latticized (see [\sureal]). The path
is required to consist only of lattice 
gauge configurations that are close to smooth continuum
gauge field configurations. (For example, this requirement
excludes deforming the lattice $g$ itself to
identity, although this is perfectly legal on 
the lattice. A single-valued phase choice
cannot be also smooth along this ``forbidden''
path.)
As a function of $t$ we expect the
non-negative [\exactlymassless] quantity
$\det{{1+V}\over 2}$ to go 
through zero a certain number
of times, $n$, which, counting multiplicities,  should
satisfy $n=2({\rm mod})4$.

The direct evaluation of 
$\det{{1+V}\over 2}$ is not
easy numerically, and even 
if it were, we would loose
some insight if I 
proceeded to establish WA that way.
Following [\odddim-\sureal] 
I choose to go about this
differently.

The fermions are understood to obey anti-periodic 
boundary conditions in all four directions 
on a four torus of equal sides
$L$. I exclude the parts of the link variables
that implement these boundary conditions from 
the link configurations quoted below. Picking
$U_{l,0}\equiv {\bf 1}$ and the $g$ from [\sureal]
I construct a curve in the 
space of Hamiltonian
matrices $H\{U_l ,\kappa\}$. 
The curve starts at
point A with $U_l=U_{l,0}$ and $\kappa=\kappa_0$
and follows $t$ from $t=0$ to $t=1$ 
at constant $\kappa$ to point B
where $U_l=U^g_{l,0}$. 
From B the curve follows
$\kappa$, at fixed gauge background, to $\kappa=0$
at point C. At $\kappa=0$ there is no dependence
on the gauge field so the curve is taken to
proceed, with constant $U_l=U_{l,0}$ this time,
increasing $\kappa$ back to $\kappa_0$ and returning
to A. Thus, we have constructed a closed path,
ABCA in the space of real matrices $H$. 
(Note that we only care about $H$ up to its 
multiplication by a positive real number;
actually, we wouldn't even care about any change 
$H\rightarrow f(H)$ with smooth, monotonic and real $f$, 
also satisfying $f(0)=0$.)
By gauge
invariance, anything that happens with $\det{{1+V}\over
2}$ along $BC$ is duplicated along $CA$, so the
number of zeros along $BCA$ 
vanishes mod four. It 
is important (and trivially true) that $\epsilon(H)$
stays well defined along the path BCA; no 
``exceptional'' configurations in the sense of the
overlap are encountered. 

Let us assume that no exceptional configurations
are encountered along the path AB either; this is
not trivial, but can be confirmed numerically. 
Then, for WA we need $\det{{1+V}\over 2}$ to vanish
$n=2({\rm mod})4$ times along AB. This happens if and
only if $|v\{ U_l ,\kappa\}>$ picks up a -1 when
transported round the loop. The nice 
thing now is that
in order to show the latter it is sufficient to establish  
that inside the disk spanned by
$\kappa ,t$ and bounded by our loop there is exactly
one double cone degeneracy point for the ground
state of ${\cal H}$ [\berry]. Note that we {\it do
not} require $V$ to be well defined 
everywhere inside 
the disk; all we need there is $H$, and $H$,
unlike $V$, is 
known to be analytic both in $\kappa$ and
in $t$. 

The degeneracy in ${\cal H}$ can come 
(generically) only from
a degeneracy in $H$ of the double cone type involving
two states closest to zero energy, each crossing zero
at the same or close-by points. The tip of the
double cone does not have to be at zero energy,
but in practice it will be reasonably close to
zero energy. A degeneracy satisfying these
requirements is excluded for $\kappa <{1\over{2d}}$
since $H$ has an impenetrable gap around zero energy
there, for any gauge field. So, we end up having
to search the interior of the region ${1\over{2d}}<
\kappa<\kappa_0,~0<t<1$ for conical degeneracies
in $H$ involving the lowest two states of $H^2$.
This is relatively easy numerically, using the
variational principle and the locality of $H$,
employing methods developed in [\ritz].
\figure{1}{\captionone}{4.0}
\figure{2}{\captiontwo}{4.0}
\figure{3}{\captionthree}{4.0}
\figure{4}{\captionfour}{4.0}
\figure{5}{\captionfive}{4.0}
\figure{6}{\captionsix}{4.0}

Working on an $L^4$ lattice with a linear interpolation
between $\{U_{l,0}\}$ and $\{U^g_{l,0}\}$ I searched
for the double cone degeneracy in $H$ and found 
exactly one
at $t=.5$ and $\kappa = .1425(5)$ for $L=8$. 
Figures 1,2,3
show three cross-sections through the double cone.
The lattice gauge
configurations were chosen with
an accidental symmetry
which ensures that 
the tip of the double cone will be 
at zero energy 
exactly. By multiplying $g$ of [\sureal] by unitary
matrices randomly drawn from a small vicinity of unity, 
the double cone is made to move slightly, 
and its
tip is no longer exactly at zero, 
as expected in the
generic case. Three cross-sections through the
double cone for the noisy gauge background at $L=8$ are shown
in figures 4,5,6. To see how the continuum limit
is approached I searched for the double cone
degeneracy for the smooth configurations on lattices
of size $L=10$ and $L=12$ and found them again at
$t=.5$ (this point has a special symmetry) but at
$\kappa=.1355(5)$ and $\kappa=.1320(5)$ respectively.
Thus, as $L\to \infty$, $\kappa$ approaches the expected
value .125 at a rate proportional to ${1\over {L^2}}$. 
The sign change in $\det(1+V)$ will occur for any $\kappa$ that
exceeds those values, at the appropriate size $L$. 

I conclude that WA is reproduced and that it
is likely that the effect will be preserved by
configurations that are not too ``rough''.

Before closing, 
I would like to make a few related comments:

(a) If $\kappa_0$ happens 
to be very close to the
degeneracy point we see 
that $\det{{1+V}\over 2}$ will
not be defined everywhere along AB; 
instead of
going through an order 2 zero, we would go through
two exceptional configurations, 
one corresponding
to the birth of an ``instanton'' and the subsequent one 
to its annihilation. A similar sequence would be
observed in the vicinity of the tip of the double
cone along a line where the gauge 
configuration is fixed
and only $\kappa$ is changing: as such
the configuration might be interpreted 
as consisting of 
an instanton and an antiinstanton, 
and would seem similar
to configurations studied in [\urs]. 
Obviously, only for quite smooth gauge configurations
and well separated lumps of non-abelian
field strength square 
and topological charge density
does the latter interpretation
become meaningful and these 
criteria do not apply
here.

(b) The double-cone degeneracies we 
find in the real
hamiltonian require the tuning of two real parameters
and thus are of the generic type. This is what makes
the phenomenon robust. The reality of H is crucial.

(c) Suppose that, following [\GW,\hasen], 
we wish to work with only 
the combination $V=\gamma_5 \epsilon (H)$ 
but not $H$ itself. $V$, unlike $H$, 
because of instantons, must
be nonanalytic in the gauge fields 
[\exactlymassless]. Thus, I see no
easy way to include the 
Berry phase mechanism as an
explanation of the sign change. 

It may seem then that
we would not even be able to understand why the
zeros required for WA 
occur in $\det{{1+V}\over 2}$
in a {\it generic} way 
(in the sense that they require tuning of
one real parameter, not two). But, this is not true: 
to understand the situation 
requires a more detailed analysis 
of the zeros
of $1+V$ than provided either in [\exactlymassless]
or in [\hasen]. Let me sketch what is needed below:

Assuming ${\rm dim} ({\rm Ker} (H))= 0$, 
define $Z$ as the subspace of the space on which $H$
acts as follows\footnote{${}^*$}
{In the more general version of the overlap, the
role of $\gamma_5$ is played by 
$\epsilon (H^\prime )$ where $H^\prime \ne 0$ and 
${\rm tr} \epsilon (H^\prime )\equiv 0$, for
all gauge fields. More explicitly, $H^\prime$
is chosen of the same form as $H$, only 
$\kappa <{1\over {2d}}$ [\npblong].}
:
$$ Z={\rm Ker}([\gamma_5 , \epsilon (H) ]).
\eqno{(3)}$$
One easily proves the following result:
$$
Z=Z_+ \oplus Z_- ,
\eqno{(4)}$$
where 
$$
Z_\pm ={\rm Ker} (1\pm V).\eqno{(5)}$$
$Z$ contains
all the eigenvectors of $V$ with real eigenvalues.
Following [\exactlymassless] we know that 
all complex eigenvalues of $V$ are paired, so
${\rm dim} (Z)$ is even. ${\rm dim} (Z_- ) +
{\rm dim} (Z_+ )$ can change only by even
numbers. If ${\rm dim} (Z_- )$ changes by an odd
amount (simplest case: by unity) this change
must be matched by a jump in ${\rm dim} (Z_+ )$.
Such a change is non-analytical since no motion
of a complex conjugate pair of eigenvalues
can smoothly yield 
a pair of $\pm 1$ eigenvalues. These ``transitions''
under variations of the background take us through
exceptional configurations where $V$ is ill defined.
The entries of the matrix $V$ are not analytic
functions of the gauge background. 
Note that at the same time that we produce
a zero eigenvector for $1+V$ we also produce one
for $1-V$, the field dependent factor of the chiral
transformation suggested in [\luscher]. (Starting
from [\npblong],
this transformation can be obtained as a remnant
of the natural chiral symmetry
fundamental to the overlap.) 

Both spaces $Z_\pm$ are invariant under $\gamma_5$.
Thus $\gamma_5$ can be diagonalized in each. The
trace of $\gamma_5$ restricted to $Z_-$ gives the
index and is exactly the deficit/surplus 
in the filling of the ground state of ${\cal
H}$ in [\npblong]. If this trace 
is non-vanishing,
obviously ${\rm dim}(Z_-) >0$, 
and the fermion
determinant vanishes. 
Note that $\det{{1-V}\over 2}$
also vanishes when the trace of  $\gamma_5$ 
restricted to $Z_+$ is non-zero. The trace of
$\gamma_5$ when restricted to $Z$, always
vanishes, since the trace of $\gamma_5$ when
restricted only to states with complex eigenvalues
of $V$ vanishes because of the pairing. 

Even if topology is trivial
in the overlap sense, there still can be states
in $Z_-$, only their total number must be even,
because the trace of $\gamma_5$ in the subspace 
has to be 
zero. A pair of such states appears at the
points where $\det{{1+V}\over 2}$ changes sign. 
Only one parameter would need to be fine tuned
to attain such a transition point rather than
the generic two: Close to this point, $V$, when
restricted to the two crossing states, is well
approximated by $-1 +A$, where $A$ is real 
and antisymmetric; a two by two antisymmetric
matrix depends on only one real parameter.
We learn that 
as long as $V$ exactly satisfies the Ginsparg-Wilson 
relation (GW)
[\GW], we could recover all
that really matters (even without knowing about Berry's phase). 
This is not surprising in view of the close 
connection between the overlap and GW emphasized in [\narayanan].
However, there is a serious 
danger that seemingly
minor truncations of $V$ would 
destroy the generic
nature of the type of level 
crossings needed for the
WA anomaly, making the levels 
avoid rather than
cross.

If all we allowed ourselves to consider 
were $V$, because of the pairing of
eigenvalues, it would be tempting to adopt
a definition of the chiral determinant as the
product of all $1+\lambda$ where $\lambda$
are the eigenvalues of $V$ with positive imaginary
part. As long as the ``fixed'' points $\pm 1$
are avoided, the definition seems fine,
even if not evidently local, at least smooth in
the background. What we showed in this paper
implies that the ``fixed'' points can't be avoided
and such a definition is questionable.
On the other hand, if we start with some fiducial
gauge background, and decide to follow the
above ``half of eigenvalues'' by analyticity in the
gauge background, we loose single-valuedness
in addition to gauge invariance. The loss of
single-valuedness becomes evident when we 
investigate a closed loop in gauge configuration
space made out of the path A to B we defined
above followed by a path B to A induced by a
smooth deformation of $g$ to unity. The existence
of this closed loop is a lattice phenomenon, and
so is the associated lack of single-valuedness.
Actually, by considering the deformations making
up the paths AB and BA simultaneously, another
disk necessarily containing a degeneracy point
of $H$ is defined and it would be interesting
to test this numerically, since this disk
is at constant $\kappa$. 

(d) The zeros of $\det(1+V)$ related to WA
occur at
zero topology and require the tuning of one real
parameter. Their role is distinct from the zeros
associated with instantons and they may be relevant
for the spontaneous breakdown of 
chiral symmetries. These
configurations might be interpreted sometimes
as instanton-antiinstantons, but, in this 
view 
the configurations are 
less distinguished than when seen
as necessary reflections of WA.
It would be useful to see these zeros by numerically
studying $1+V$ directly. For the gauge
configurations studied here we know where to look
for these zeros: for example, in the smooth case,
on an $8^4$ lattice, they should occur 
for $t =.5$ and any $\kappa > .143$. 

(e) Any continuum argument based on 
topology uses gauge
backgrounds that are atypical in the path integral
in the sense that they are too smooth. Such smooth
configurations have most likely zero probability in
the path integral. By taking the effect to the lattice
we open the way to directly check whether the ``noisy''
backgrounds which are typical to the path integral
preserve the features found assuming smoothness. 
All continuum considerations I am aware of have
weathered very well this ``latticization". Still,
I feel that at a strict level, something conceptual
is being gained by going to the lattice in the way
done here.

(f) The Wigner-Brillouin phase choice (WB)
[\npblong] will introduce
new exceptional configurations where it is ill defined.
The existence of these new exceptional points is 
necessary since the phase choice attempts to defined
a section in some twisted $Z_2$-bundles
over certain submanifolds in the space
of parameters of the Hamiltonians $H$. One would
need to know where these new exceptional points are
to determine to what extent the WB phase choice
reproduces the WA. In [\odddim] we saw that the WB
phase choice often does reproduce the required global
anomaly, but not always.

(g) The gauge transformation $g$ can be deformed to unity
on the lattice. 
For the WA to work correctly a WB
exceptional point is needed also on the path connecting 
B to A by $g$-deformation to accommodate the needed
jump in sign. 

(h) Nontrivial anomaly cancelation in 
this case could 
be seen for example by taking a single Weyl multiplet
with $I={3\over 2}$ [\sureal]. There should be no
odd numbers of double-cone degeneracies of the type seen above.
Note that unlike in the case of nontrivial cancelation of
perturbative anomalies [\geom],
there is no Berry curvature over the space of gauge configurations
that needs to be eliminated before 
a ``perfect'' (in the sense of
[\npblong], unrelated to [\hasen]) regularizations of the chiral
gauge theory in question should become possible.

Another potential 
example of a chiral theory with similar
properties would be a model containing one $I={1\over 2}$
Weyl fermion and another $I={5\over 2}$ Weyl fermion.

\vskip .5cm 
This work was supported in 
part by the DOE under 
grant \#DE-FG05-96ER40559.
I have benefited from discussions with Y. Kikukawa
about global anomalies and with 
R. Narayanan about 
references [\ritz,\urs,\narayanan,\luscher]. 
I am grateful to Claudio Rebbi
for providing me with the 
qcdf90 package [\rebbi]
which I
adapted to $SU(2)$ with a real Wilson-Dirac operator [\sureal]
and used in the numerical part of the above work.

\vfill\eject
\centerline{\bf References.}
\medskip 
\item{[\witten]} E. Witten, Phys. Lett. B117
(1982) 324.
\item{[\elitzur]} S. Elitzur, V. Nair, Nucl. Phys. 
B243 (1984) 205; F. R. Klinkhamer, Phys. Lett. B256
(1991) 41.
\item{[\gaumenelson]} P. Nelson, L. Alvarez-Gaume,
Comm. Math. Phys. 99 (1985) 103.
\item{[\npblong]} R. Narayanan, H. 
Neuberger, Nucl. Phys. B 443 (1995) 305.
\item{[\daemi]} S. Randjbar-Daemi, J. Strathdee, 
Phys. Lett. B348 (1995) 543.
\item{[\kaplanschmaltz]} D. Kaplan, M. Schmaltz,
Phys.Lett. B368 (1996) 44.
\item{[\odddim]} Y. Kikukawa, H. Neuberger, 
Nucl.Phys. B513 (1998) 735.
\item{[\geom]} H. Neuberger, hep-lat/9802033.
\item{[\sureal]} H. Neuberger, hep-lat/9803011.
\item{[\exactlymassless]} H. Neuberger, 
Phys.Lett. B417 (1998) 141.
\item{[\berry]} G. Herzberg, H. C. Longuet-Higgins,
Disc. Farad. Soc. 35 (1963) 77. For more details
see the original contribution by M. V. Berry titled
``The Quantum Phase, Five Years After'', in the book
``Geometric Phases in Physics'' by A. Shapere and F. Wilczek,
(Worl Scientific, 1989). 
\item{[\ritz]} B. Bunk, K. Jansen, M. 
L{\" u}scher, H. Simma, DESY-Report(Sept. 94);
T. Kalkreuter, H. Simma, Comp. Phys. Comm. 93 (1996) 33.
\item{[\urs]} R. Edwards, U. Heller, R. Narayanan,
hep-lat/9801015.
\item{[\GW]} P. Ginsparg, K. Wilson, Phys. Rev. D25
(1982) 2649.
\item{[\hasen]} P. Hasenfratz, V. Laliena,
F. Niedermayer, hep-lat/9801021.
\item{[\narayanan]} R. Narayanan, hep-lat/9802018.
\item{[\luscher]} M. L{\" u}scher, hep-lat/9802011.
\item{[\rebbi]} I. Dasgupta, A. R. Levi, V. Lubicz, 
C. Rebbi, Comp. Phys. Comm. 98 (1996) 365.
\vfill\eject 
\end